\documentclass[conference]{IEEEtran}
\usepackage{blindtext, graphicx}
 \usepackage{multirow}
 \usepackage[table,xcdraw]{xcolor}
 \usepackage[bookmarks=false]{hyperref}

\usepackage[noadjust]{cite}

\makeatletter

\ifCLASSINFOpdf
\else
\fi
\renewcommand\IEEEkeywordsname{Index Terms}

\begin{document}

%
\title{Feasibility Study of OFDM-MFSK Modulation Scheme for Smart Metering Technology}

\author{\IEEEauthorblockN{Ghaith Al-Juboori, Angela Doufexi and Andrew R. Nix}
\IEEEauthorblockN{Communication Systems and Networks Group-Department of Electrical and Electronic Engineering\\University of Bristol, Bristol, United Kingdom. \\Email: {Ghaith.al-juboori, a.doufexi, Andy.nix}@bristol.ac.uk
}}


%


\maketitle

\begin{abstract}
The Orthogonal Frequency Division Multiplexing based M-ary Frequency Shift Keying (OFDM-MFSK) is a non-coherent modulation scheme which merges MFSK with the OFDM waveform. It is designed to improve the receiver sensitivity in the hard environments where channel estimation is very difficult to perform. In this paper, the OFDM-MFSK is suggested for the smart metering technology and its performance is measured and compared with the ordinary OFDM-BPSK. Our results show that, depending on the MFSK size value (\textbf{\textit{M}}), the Packet Error Rate (PER) has dramatically improved for OFDM-MFSK. Additionally, the adaptive OFDM-MFSK, which selects the best \textbf{\textit{M}} value that gives the minimum PER and higher throughput for each Smart Meter (SM), has better coverage than OFDM-BPSK. Although its throughput and capacity are lower than OFDM-BPSK, the connected SMs per sector are higher. Based on the smart metering technology requirements which imply the need for high coverage and low amount of data exchanged between the network and the SMs, The OFDM-MFSK can be efficiently used in this technology.
\end{abstract}


%
\IEEEpeerreviewmaketitle

\section{Introduction}
The smart metering is one of the significant technologies that will be used to effectively manage energy systems in the future. This technology will provide new information and services for both energy companies and consumers which lead to reduce costs and carbon emissions. By 2020, the number of installed smart meters (electricity, gas, and water) is projected to rise to 1.6 billion \cite{RN78}. In general, the smart meter (SM) is defined as an element of two-way communication between the domestic meter and the utility provider to effectively gather details energy usage information \cite{RN61}.

The radio coverage for this technology represents an essential consideration due to installing these meters in challenging communication environments and also the need for getting near 100\% coverage. Moreover, low cost and low power consumption smart metering devices also represent significant requirements for this technology. Additionally, the amount of the exchanged data between the SMs and the network is relatively low and can be classified to fall into the category of Internet of Things (IoT) applications \cite{RN69}. Many studies, such as \cite{RN69} \& \cite{RN59}, studied different available techniques and suggested a certain solution for this technology.

The non-coherent detected $M$-ary  Frequency Shift Keying (MFSK) in conjugation with Orthogonal Frequency Division Multiplexing (OFDM) waveform (OFDM-MFSK) was suggested as a robust transmission technique in the hard environments as the fast fading channels and high-speed applications such as high-speed trains \cite{RN52}. This method does not need equalisation and channel estimation processes; this leads to a very simple \& low-cost receiver structure. Furthermore, the OFDM-MFSK technique gives a high receiver sensitivity, as it is illustrated in section-II.

In this paper, we studied the ability to apply OFDM-MFSK as a solution for the smart metering technology and compared its performance with the ordinary OFDM-BPSK in different cases and scenarios. The comparison includes the Packet Error Rate (PER), throughput, coverage and capacity performance for both of them. The remainder of this paper is sorted as follows: section-II gives a brief description of the OFDM-MFSK technique. Details about the modelling approach, assumptions and the channel model are provided in section-III. In section-IV, the performance analysis and results are shown for both modulation techniques. Finally, conclusions are drawn in the section-V.

\section{OFDM-MFSK Overview}

The MFSK is a famous modulation scheme which is used to get the robust transmission in the hard environments. The OFDM-MFSK is an integration between OFDM and MFSK which allows to group $M$ sub-carriers into a subset and applies MFSK to each one of these subsets (groups). The non-coherent detection is permitted in this modulation scheme which is needed for many scenarios where no channel estimation is required such as fast fading environments \cite{RN52}.

The basic concept of the OFDM-MFSK modulation scheme, using $M$=4, is shown in Fig. 1. For simplicity, MFSK \& BPSK are used to refer for OFDM-MFSK \& OFDM-BPSK respectively in the remainder of this paper. Each group of sub-carriers, four in this case, are gathering into a subset. In each subgroup, only one sub-carrier is chosen for transmission whereas no energy is transmitted on the other sub-carriers. The selection of the active sub-carrier in each subset depends on the data bits. As illustrated in Fig.1, $log_{2}(M)$ bits, 2 bits in this case, are allocated for each subset using Grey code.

\begin{figure}[hbpt!]
\centering
\includegraphics[width=3.5in]{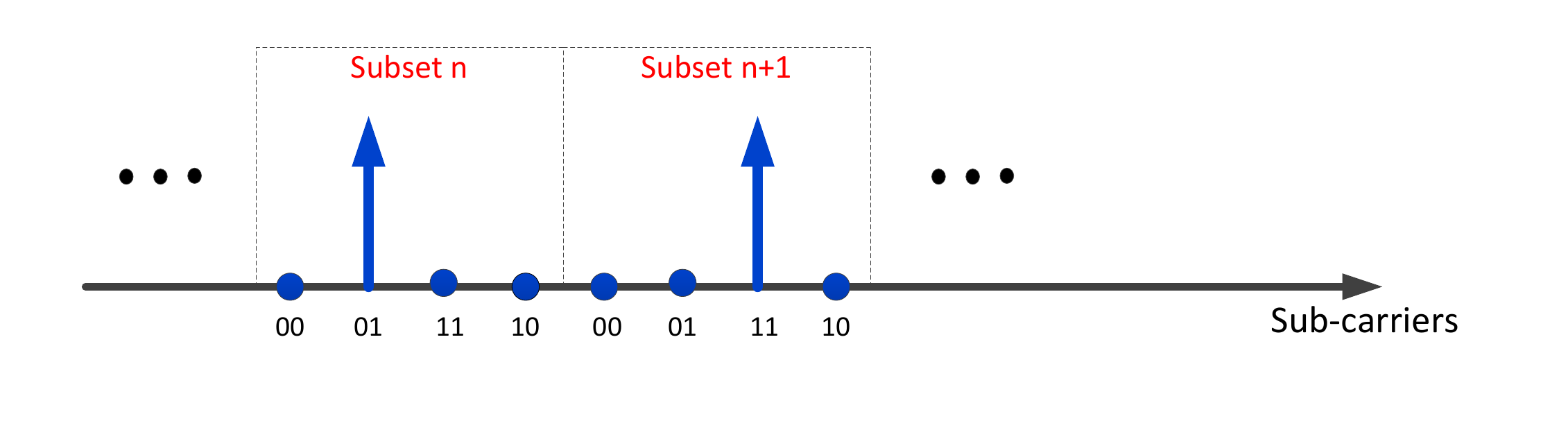}
\caption{Basic concept of the OFDM-4FSK modulation scheme.}
\label{fig_1}
\end{figure}

This modulation scheme is engineered to improve receivers sensitivity; however, this improvement is at the expense of the bandwidth efficiency. In MFSK, the higher the value of $M$, the better the receiver sensitivity but at the cost of lower spectral efficiency. the MFSK bandwidth utilisation equals to $log_{2}(M)/M$ whereas it equals to $log_{2}(M)$ for M-QAM \cite{RN52}. However, this reduces the spectral efficiency, and this represents the main disadvantage of MFSK.

Several methods were proposed to tackle this issue such as the hybrid transmission method where additional data can be sent by exploiting the phase of the occupied sub-carriers. This is done by combining MFSK (OFDM-MFSK) with the Differential Phase Shift Keying (DPSK) \cite{RN50}. This combination is allowed because MFSK, non-coherent detection scheme, permits a random phase selection for all the occupied sub-carriers in the transmitter. Additionally, an another method to exploit this degree of freedom (the random phase of the occupied sub-carriers) to reduce the Peak-to-average power ratio (PAPR) was proposed in \cite{RN51}.

Channel coding coupling with an interleaver is used to mitigate the channel effects such the frequency selective effect which can lead to entirely fade some sub-carriers and produce an error floor. To get the best performance, the soft decision detection is used to provide a degree of reliability for each bit to the decoder. An appropriate log-likelihood metric for the $n^{th}$ bit of a coded symbol in a transmission is calculated, as follows, based on the components of the received vector $r_{i}$:
\begin{equation}
    L_{n}=max_{i\in S_n^1}|r_{i}|^{2}-max_{i\in S_n^0}|r_{i}|^{2} .
\end{equation}
$S_n^0$ is the subset of all components indicators where the code symbols have ''0'' at the $n^{th}$ digit of the bit mapping. Accordingly, there is a ''1'' at the $n^{th}$ digit of the bit mapping in the other case ($S_n^1$) \cite{RN50}.

\section{Simulation Approach}
\subsection{Modelling Approach and Assumptions}
The block diagram of the approach used for modelling MFSK \& BPSK based on the LTE-A like parameters for smart metering applications is shown in Fig 2. The main two components are the coverage and capacity analysis since we focus on coverage and capacity in the comparison between MFSK and BPSK modulation schemes. The coverage analysis estimates the max coverage radius and the outage probability based on the parameters of the modulation type and channel propagation model. On the other hand, the capacity analysis estimates the aggregate throughput of a sector and also the average capacity per SM based on the density of these SMs and LTE parameters. Based on the deployment environment (urban \& rural), the channel propagation model is calculated and used for both analyses, as illustrates in the following sub-section.

\begin{figure}[hbpt!]
\centering
\includegraphics[trim=15mm 0mm 15mm 0mm,width=3.5in]{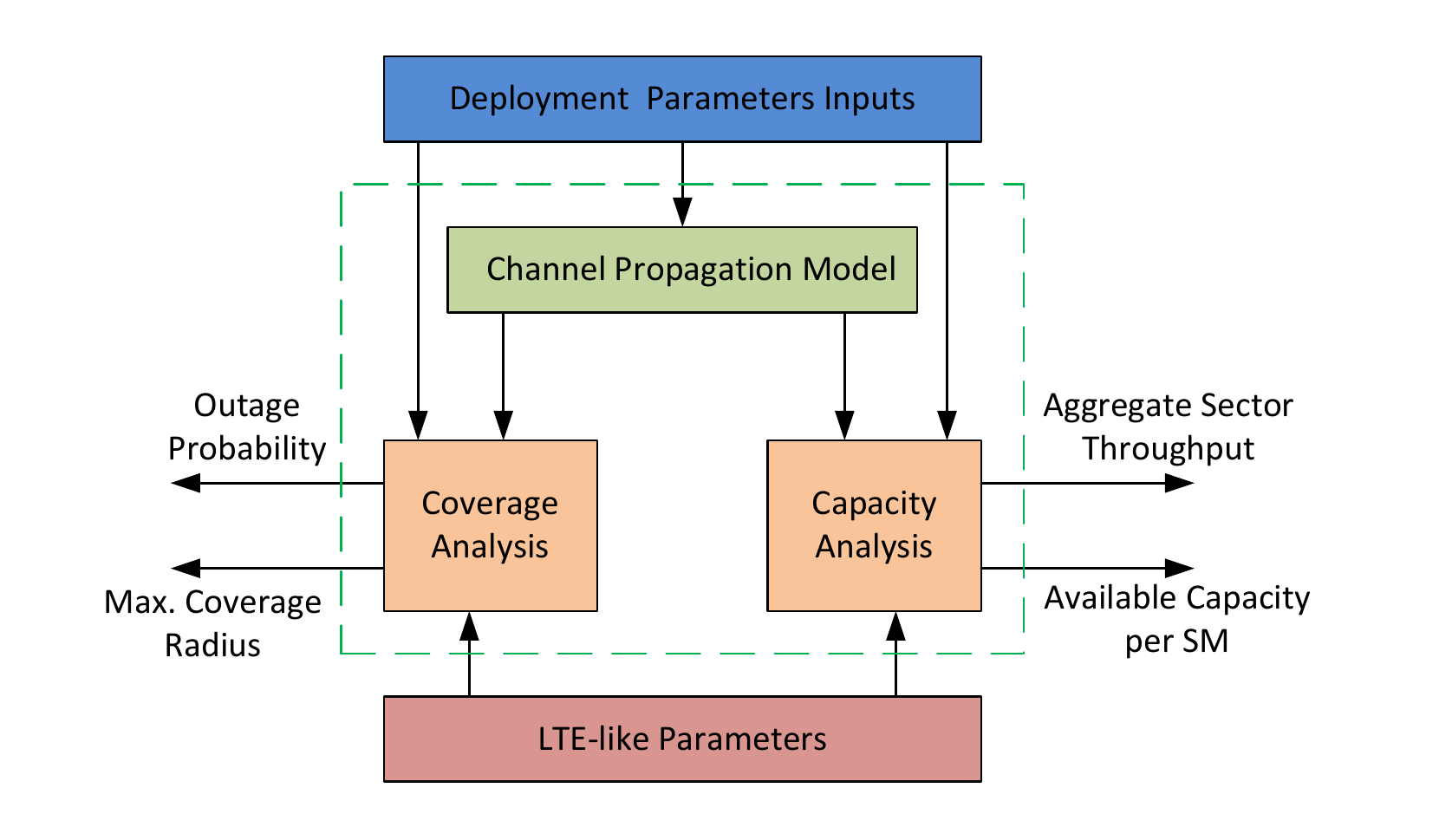}
\caption{Block diagram of the modelling approach.}
\label{fig_2}
\end{figure}

The 3GPP macro-cellular deployment with unity frequency reuse factor is performed. There are three sectors in each cell with cell radius, diameter and Inter-Site Distance (ISD) equal to R, 2R \& 3R respectively \cite{RN28}. The SMs are randomly and uniformly distributed in the cell at a distance between 50 m and the max cell diameter from the BS. An operating frequency of 900 MHz and a bandwidth of 3MHz were assumed. The main parameters in this study are listed in Table-\ref{Table-I}.

\begin{table}
\centering
\caption{Simulation Parameters.}
\label{Table-I}
\begin{tabular}{|l|l|l|}
\hline
\rowcolor[HTML]{9B9B9B} 
\multicolumn{2}{|l|}{\cellcolor[HTML]{9B9B9B}\textbf{Parameter}} & \textbf{Value} \\ \hline
 & Transmission Power (DL) & 32dBm \\ \cline{2-3} 
 & Peak Antenna Gain & 12 dBi \\ \cline{2-3} 
 & Noise Figure & 5 dB \\ \cline{2-3} 
\multirow{-4}{*}{Base Station (BS)} & Antenna Type & As it is mentioned in \cite{RN28} \\ \hline
 & Transmission Power (UL) & 24 dBm \\ \cline{2-3} 
 & Antenna Gain & 0 dBi \\ \cline{2-3} 
 & Noise Figure & 9 dBm \\ \cline{2-3} 
\multirow{-4}{*}{\begin{tabular}[c]{@{}l@{}}Smart Meter \\ (SM)\end{tabular}} & Antenna Type & Omnidirectional \\ \hline
 & Uplink (UL) & 3 MHz \\ \cline{2-3} 
\multirow{-2}{*}{Bandwidth} & Downlink (DL) & 3 MHz \\ \hline
 & Urban-Macro & R=250, 500, 750, 1000 m \\ \cline{2-3} 
\multirow{-2}{*}{\begin{tabular}[c]{@{}l@{}}Environment \&\\ Cell Radius\end{tabular}} & Rural-Macro & R=2, 4, 6, 8, 10 km \\ \hline
\multicolumn{2}{|l|}{Carrier Frequency} & 900 MHz \\ \hline
\multicolumn{2}{|l|}{BS-SM distance} & 50-Max cell diameter. \\ \hline
 & MFSK & M=2, 4, 8, 16, 64, 256 \\ \cline{2-3} 
\multirow{-2}{*}{Modulation Scheme} & BPSK & ordinary BPSK \\ \hline
\multicolumn{2}{|l|}{Channel Coding} & LDPC \cite{RN54} \\ \hline
\multicolumn{2}{|l|}{Coding Rate} & 1/2 \\ \hline
\multicolumn{2}{|l|}{Input Data Block Size} & 204 bits \\ \hline
\multicolumn{2}{|l|}{OFDM Symbol Size} & 256 \\ \hline
\multicolumn{2}{|l|}{Cyclic Prefix} & 32 \\ \hline
\multicolumn{2}{|l|}{Number of the SMs per cell (K)} & \begin{tabular}[c]{@{}l@{}}Depends on the SMs' \\ density and the cell's radius\\ (See (7))\end{tabular} \\ \hline
\end{tabular}
\end{table}

\subsection{Channel Model}
Fig. 3 illustrates the end to end radio link in the smart metering system. It is clear that a signal incurs different fading and losses during its travel in the different environments. The total losses ($L_{total}$) for each link can be expressed as follows:
\begin{equation}
    L_{total}=L_{outdoor_{-}losses}+L_{penetration}+L_{indoor_{-}losses}.
\end{equation}
Based on the channel propagation model in \cite{RN28}, the outdoor losses, $L_{outdoor_{-}losses}$ in dB, with a distance d (in km) can be modelled as:
\begin{equation}
    L_{outdoor_{-}losses}=L_{o}+10\acute{n}log_{10}(d)+X,
\end{equation}
where $L_{o}$ \& $\acute{n}$ are the path loss reference and exponent respectively, and their values depend on the environment as shown in the Table-\ref{Table-II}. X represents the shadowing loss which can be expressed as a log-Normal distribution variable with a standard deviation of 10 dB \cite{RN66}.

\begin{figure}[hbpt!]
\centering
\includegraphics[width=3.5in]{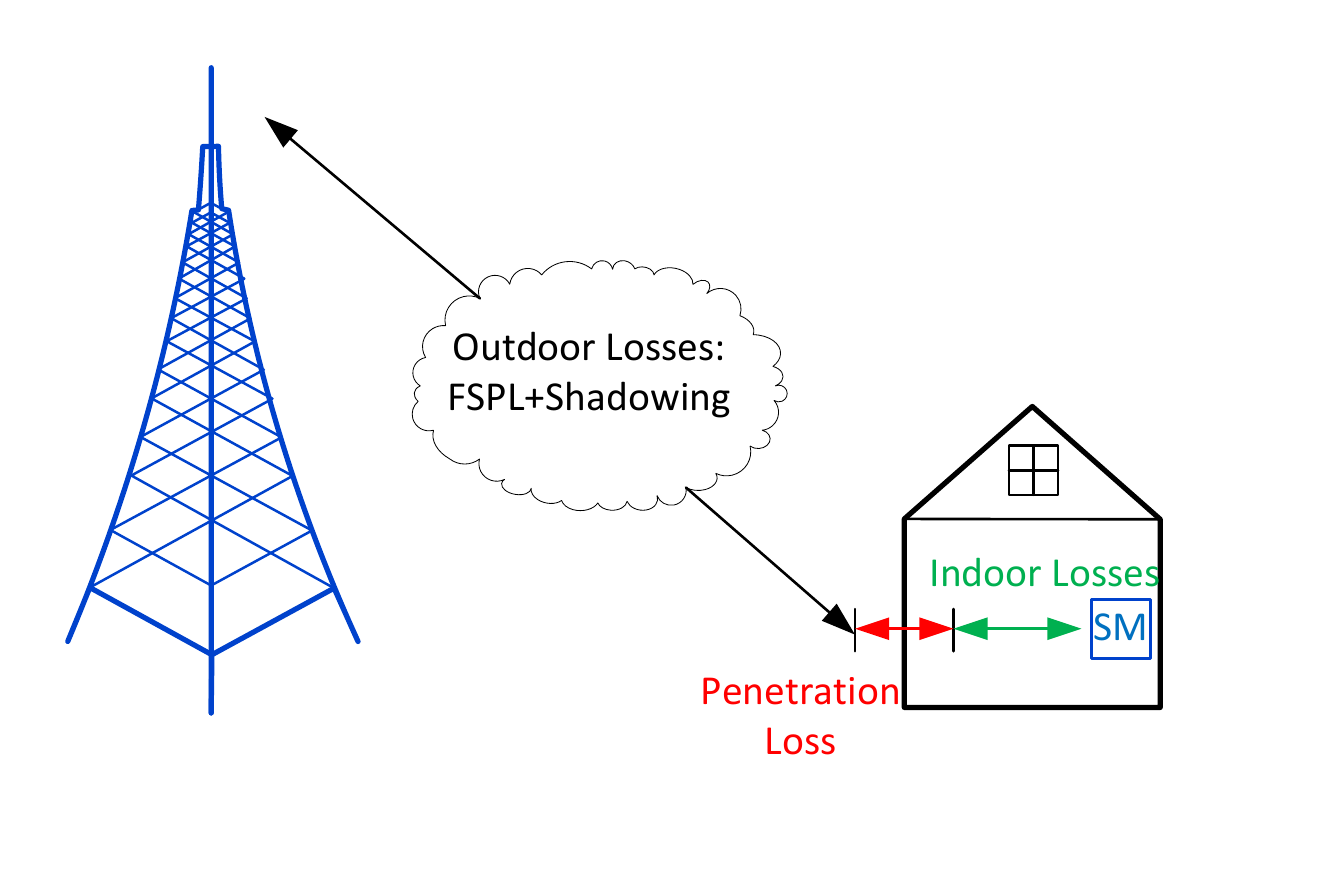}
\caption{End to end radio link losses in smart metering system.}
\label{fig_3}
\end{figure}

In this study, the penetration loss $L_{penetration}$ and indoor loss $L_{indoor_{-}loss}$ are chosen to be 12 \& 8 dB respectively \cite{RN59}. Additionally, only one wall is assumed to exist in each link between the BS and SM.
\begin{table}[hbpt!]
\centering
\caption{Path Loss Parameters at 900 MHz.}
\label{Table-II}
\begin{tabular}{|l|l|l|}
\hline
\rowcolor[HTML]{9B9B9B} 
Environment & $L_{o}$ & $\acute{n}$ \\ \hline
Urban Area & 120.9 & 3.76 \\ \hline
Rural Area & 95.5 & 3.41 \\ \hline
\end{tabular}
\end{table}

Based on the path loss and device parameters, the received power and the signal to noise ratio (SNR) can be calculated as:
\begin{equation}
    P_{rx}=P_{tx}+G_{tx}+A_{rad}-L_{total}+G_{rx}.
\end{equation}
$P_{tx}$ represents the transmit power in dBm, $G_{tx}$ and $G_{rx}$ are the transmit and received antenna gains in dBi, and $A_{rad}$ is the BS antenna radiation pattern in dB as shown in \cite{RN28}. The SNR (in dB), can be expressed as:
\begin{equation}
    SNR=P_{rx}-P_{N},
\end{equation}
where $P_{N}$ is the noise power in dBm and it can be expressed as:
\begin{equation}
    P_{N}=-198.6+10log_{10}(BT)+F,
\end{equation}
where $B$ is the bandwidth, $T$ is the temperature in Kelvin, and $F$ is the device noise figure.

The number of SMs in each sector is determined as follows:
\begin{equation}
    No. of SMs=\rho\pi R^{2} .
\end{equation}
 $R$ is the cell radius in km, and $\rho$ is the SMs' density (i.e., the No. of SMs per square km), and it's equal to 2000 $SM/km^{2}$ and 10 $SM/km^{2}$ in the urban and rural scenarios respectively \cite{RN66}.

\section{Results and Analysis}
\subsection{Performance Comparison in AWGN Channel}
Fig. 4 illustrates the PER versus SNR performance for MFSK with different $M$ values and the BPSK in AWGN channel, for more information about parameters refer to Table-\ref{Table-I}. As it is seen, the MFSK performance overcomes BPSK especially at high values of $M$ ($M\geq8$) while the MFSK performance becomes worse as $M$ decreases ($M\leq4$). Moreover, a remarkable SNR gain, between 1.7 to 14 dB, can be seen in the MFSK modulation scheme with $M\geq8$ compared to BPSK (in Rayleigh channel the difference is between 0-11 dB see \cite{RN80}). This gain will lead to significant improvements in the smart meter applications, as it will be seen in the following sub-sections. This graph is also used to determine the threshold SNR values required to achieve PER levels equal to $1\times 10^{-3}$  for both modulation schemes in this paper.

\begin{figure}[hbpt!]
\centering
\includegraphics[trim=25mm 90mm 25mm 90mm,width=3.5in]{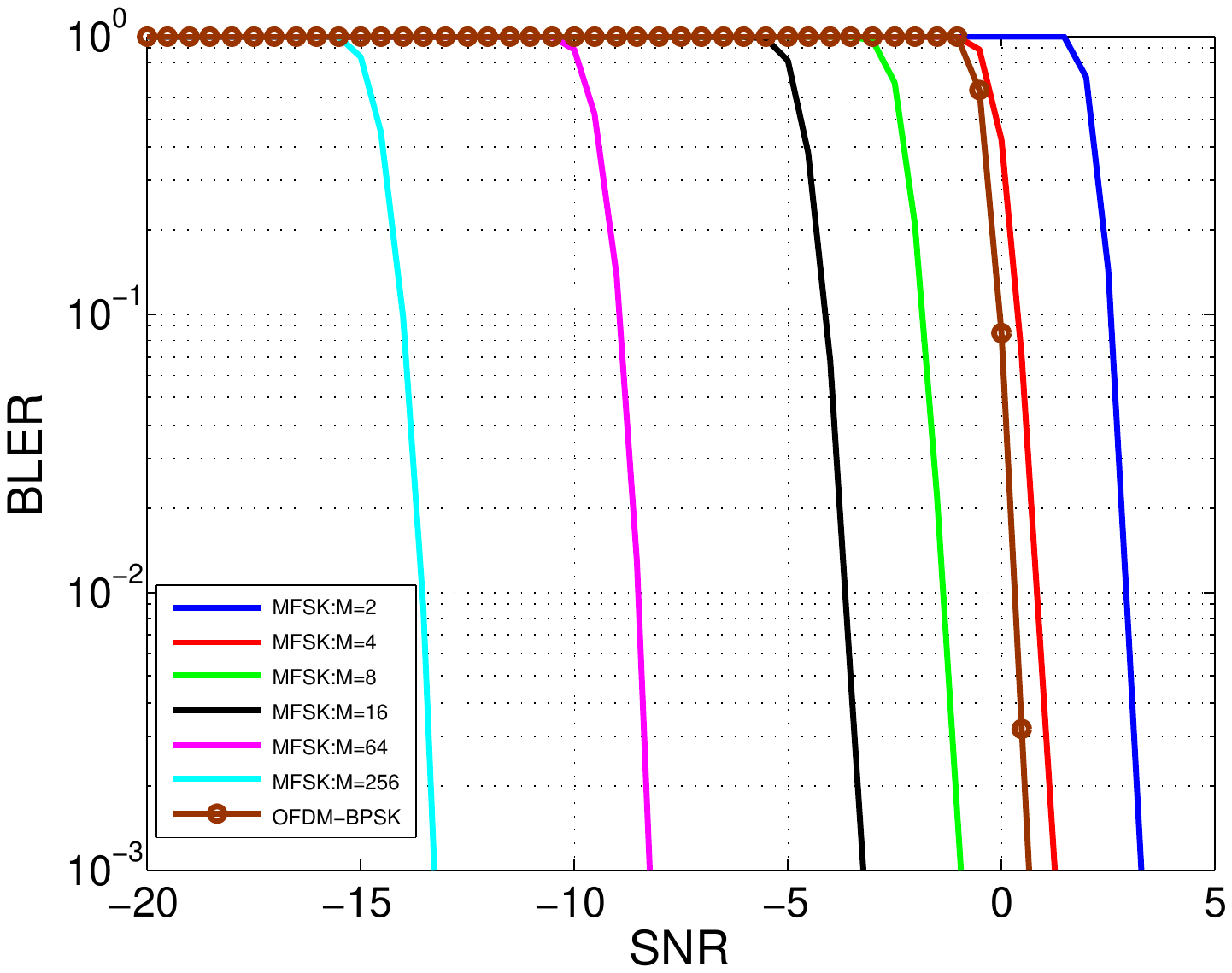}
\caption{PER performance for MFSK (with different $M$) and BPSK in AWGN channel.}
\label{fig_4}
\end{figure}

\subsection{System Level Performance in Urban Uplink Scenario}
In this subsection, the system level performance of the MFSK modulation scheme with different $M$ values is measured and compared with the BPSK performance in term of PER and throughput in an urban uplink scenario with cell radius equals to 500m, using the channel model as explained in section III. Fig. 5 depicts the Cumulative Distribution Function (CDF) of the SMs' SNR in this case. This figure shows the need to a robust communication scheme in this technology to achieve a good coverage due to the fact that 6\% of the SMs have SNRs less than -10 dB and around 25\% of them have SNRs less than 0 dB.

\begin{figure}[hbpt!]
\centering
\includegraphics[trim=25mm 90mm 25mm 90mm,width=3.5in]{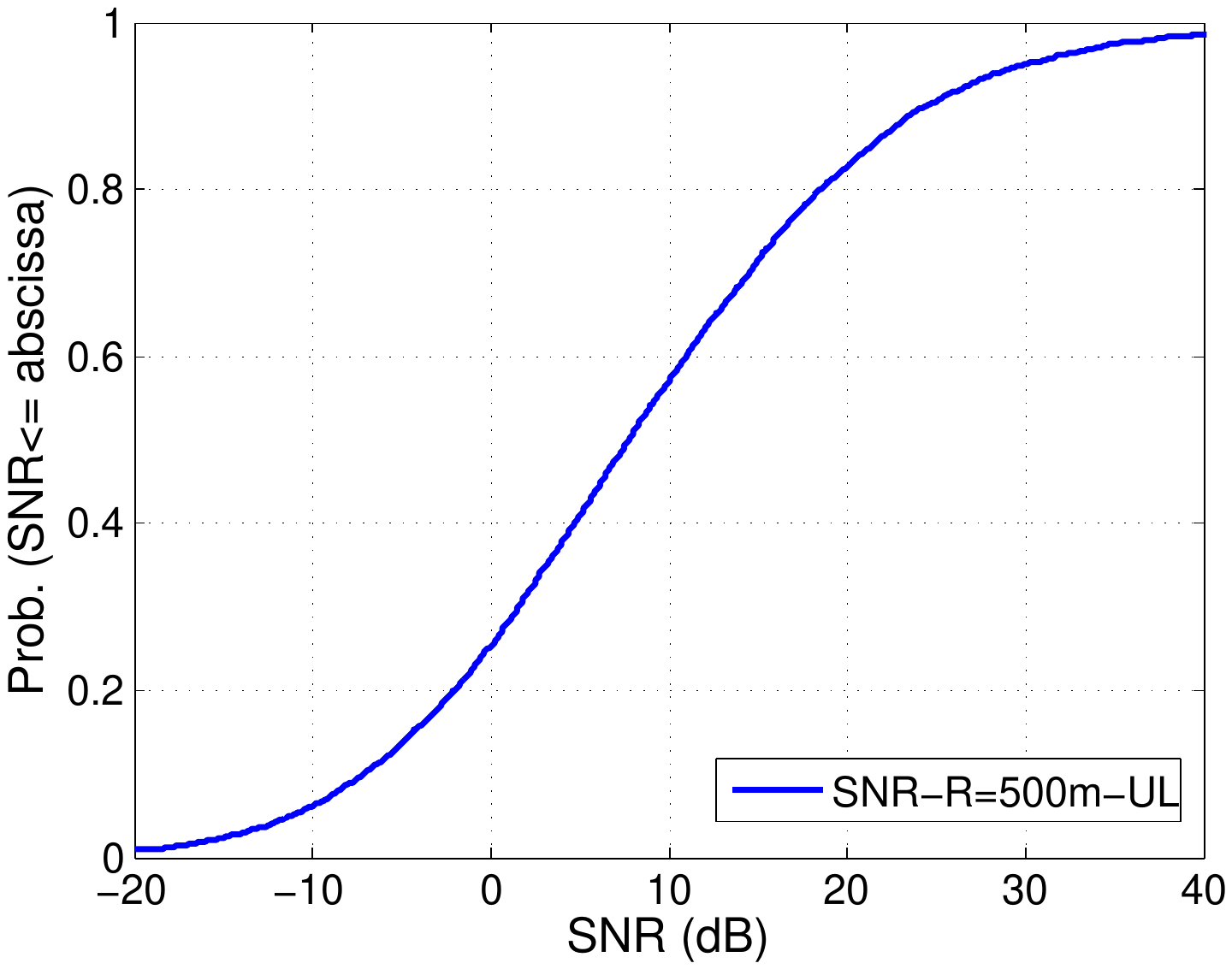}
\caption{CDF of the SMs' SNR for uplink urban scenario R=500m.}
\label{fig_5}
\end{figure}

\begin{figure}[hbpt!]
\centering
\includegraphics[trim=25mm 90mm 25mm 90mm,width=3.5in]{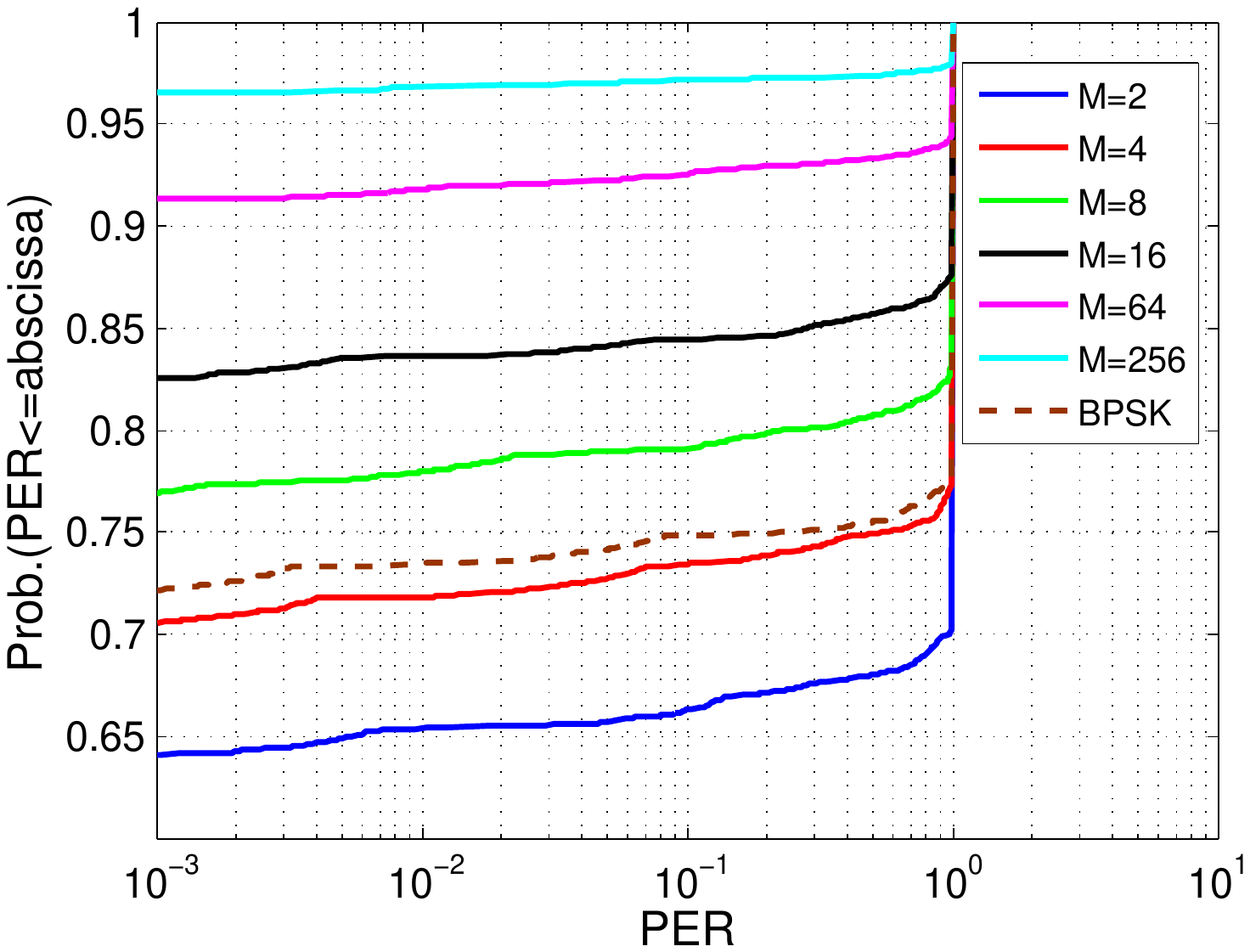}
\caption{CDF of the SMs' PER with MFSK \& BPSK modulation schemes.}
\label{fig_6}
\end{figure}
Fig. 6 shows the CDF of the PER for the SMs in this case. It is clear that more than 95\% of the SMs, in the MFSK-$M=256$ case, have PER values less than or equal to $1 \times 10^{-3}$ and this percentage decreases as $M$ declines to reach just below 65\% in case $M=2$. On the other hand, in the case of ordinary BPSK, around 72\% of the SMs have this PER value. Also, it is interesting to note that MFSK with $M\geq16$ has a remarkable PER difference when comparing with BPSK. 

\begin{figure}[hbpt!]
\centering
\includegraphics[trim=25mm 90mm 25mm 90mm,width=3.5in]{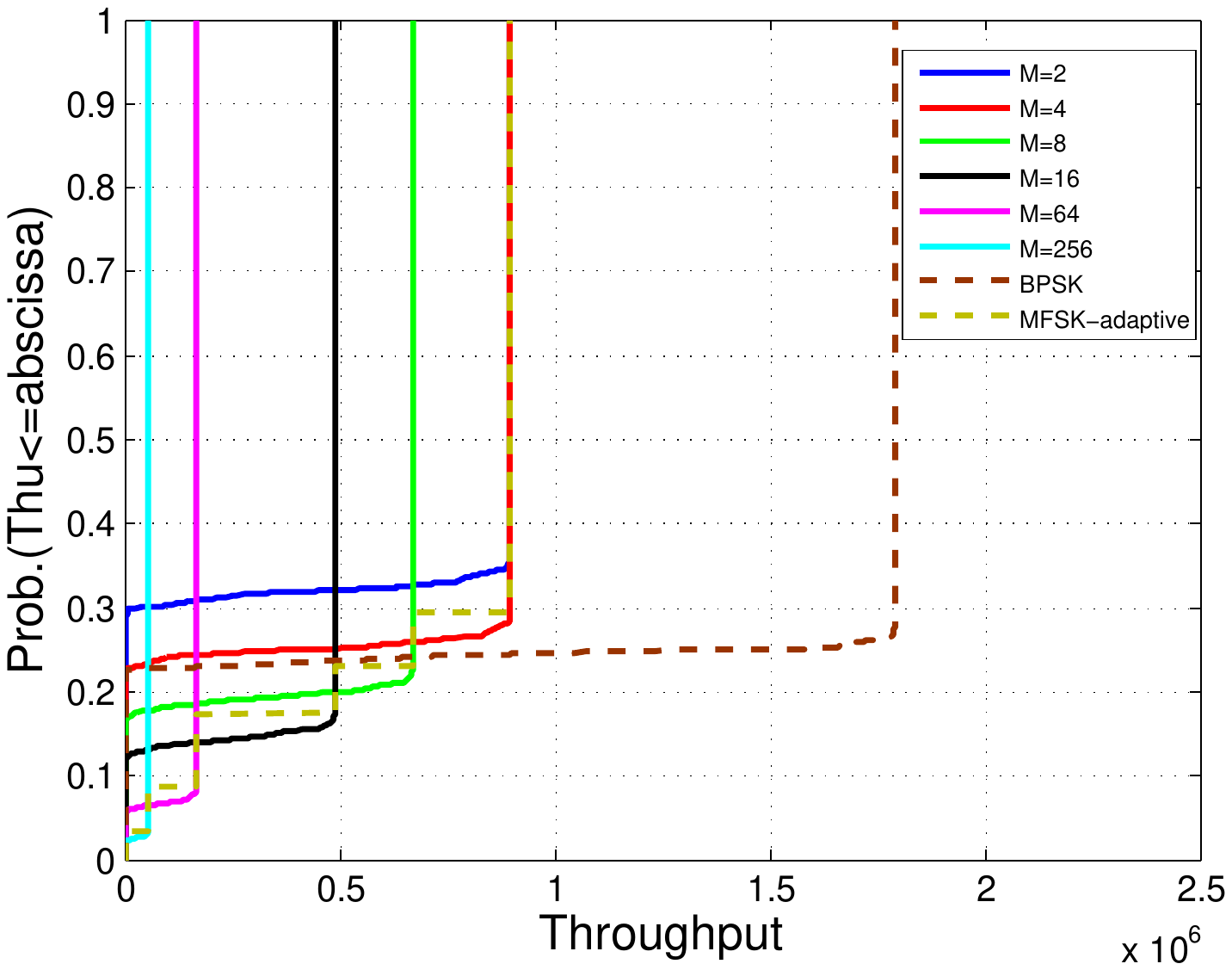}
\caption{CDF of the SMs' throughput with different modulation schemes.}
\label{fig_7}
\end{figure}

Fig. 7 shows the CDF of the throughput for the SMs in this case. We can observe that around 97\% of the SMs have the max throughput in case of MFSK-$M$ =256 and this percentage value decreases as M decreases. Additionally, in the MFSK case, it is clear that maximum throughput decreases with increase M. The maximum throughputs are (892, 892, 672, 448, 168, 56) Kbps for $M$=2, 4 , 8, 16, 64, 256 respectively whereas it equals to 1.792 Mbps in the BPSK case. These results agree with the fact that the MFSK modulation scheme is designed to improve the receiver sensitivity and this improvement is at the cost of the bandwidth efficiency. Additionally, Fig. 7 also depicts the CDF of the throughput of the adaptive MFSK scheme in which for each SM the lowest value for $M$ that permits PER level less than or equal to $1 \times 10^{-3}$ is chosen. The adaptive MFSK is already used when comparisons with BPSK in coverage and capacity analyses are performed. 

\subsection{Coverage Analysis}
The coverage analysis is used to determine the maximum cell diameter (or radius) that satisfies a particular performance criterion, such as maximum outage probability. In this study, as in \cite{RN66}, the applied coverage criterion is that the median SNR in the uplink case is equal or greater than that desired by the lowest most robust case, in the adaptive MFSK it represents MFSK with $M=256$. The choice of the uplink case because it has less transmission power compared to the downlink case which leads to more limiting. The maximum cell diameter can be written as:
\begin{equation}
    D_{max}=max\left\{ d:\overline{SNR(d)}\geq\gamma_{o}\right\},
\end{equation}
where $\gamma_{o}$ is the minimum desired SNR to achieve a PER of $1\times10^{-3}$ when using the lowest most robust MCS.
Based on the threshold values for MFSK with different $M$ and BPSK which were obtained from Fig. 4, Table-\ref{Table-III} illustrates the max cell diameter for each case. It is clear that the cell coverage in the rural environment is larger than that in the urban environment, the reason for that is the lower losses in the first scenario compared to the second.
 
\begin{table}
\centering
\caption{Max Cell Diameter in km.}
\label{Table-III}
\begin{tabular}{|l|l|l|l|l|l|l|l|}
\hline
\cellcolor[HTML]{9B9B9B} & \multicolumn{6}{c|}{\cellcolor[HTML]{9B9B9B}MFSK} & \cellcolor[HTML]{9B9B9B} \\ \cline{2-7}
\multirow{-2}{*}{\cellcolor[HTML]{9B9B9B}\begin{tabular}[c]{@{}l@{}}variable or \\ Environ. Type\end{tabular}} & M=256 & M=64 & M=16 & M=8 & M=4 & M=2 & \multirow{-2}{*}{\cellcolor[HTML]{9B9B9B}BPSK} \\ \hline
Thresh. SNR & -13.25 & -8.25 & -3.25 & -0.75 & 1.25 & 2.9 & 0.75 \\ \hline
Urban Environ. & 1.85 & 1.36 & 1 & 0.85 & 0.76 & 0.69 & 0.78 \\ \hline
Rural Environ. & 10.94 & 7.8 & 5.57 & 4.7 & 4.11 & 3.68 & 4.25 \\ \hline
\end{tabular}
\end{table}

The outage probability represents a substantial factor for the performance assessment of the wireless systems, and it measures the failing probability to achieve a specified SNR value required for a particular service, it can be expressed as \cite{RN68}:
\begin{equation}
    Pr._{outage}=Pr[SNR\leq\gamma_{o}]
\end{equation}

Fig. 8 \& 9  show the outage probability for the urban and rural environments respectively. It is interesting to note that the adaptive MFSK has lower outage probability (higher coverage) than BPSK for uplink and downlink in both environments. Based on the traditional coverage network condition which allows to only 5\% outage probability, the adaptive MFSK doubles the coverage from 300 m, in the BPSK, to 600 m. Furthermore, depend on the above condition in the rural environment, the adaptive MFSK has a coverage of 3 km, whereas BPSK has a coverage of 0.75 km. This means that MFSK has four times higher coverage than BPSK. Moreover, the results show that MFSK can be used in the ultra reliable applications where 99.99\% coverage condition need to be achieved in the urban scenario with a cell radius of 350 m, while BPSK fails to achieve this condition. Additionally, the difference between the downlink \& uplink cases increases with increase the cell radius in both environments. This happens due to increase the losses as increase the distance in both cases while the transmit power is higher in the downlink case. Finally, for the same value of the outage probability, the rural environment has a coverage range that far exceeds the urban due to its lower losses.
 
\begin{figure}[hbpt!]
\centering
\includegraphics[trim=25mm 90mm 25mm 90mm,width=3.5in]{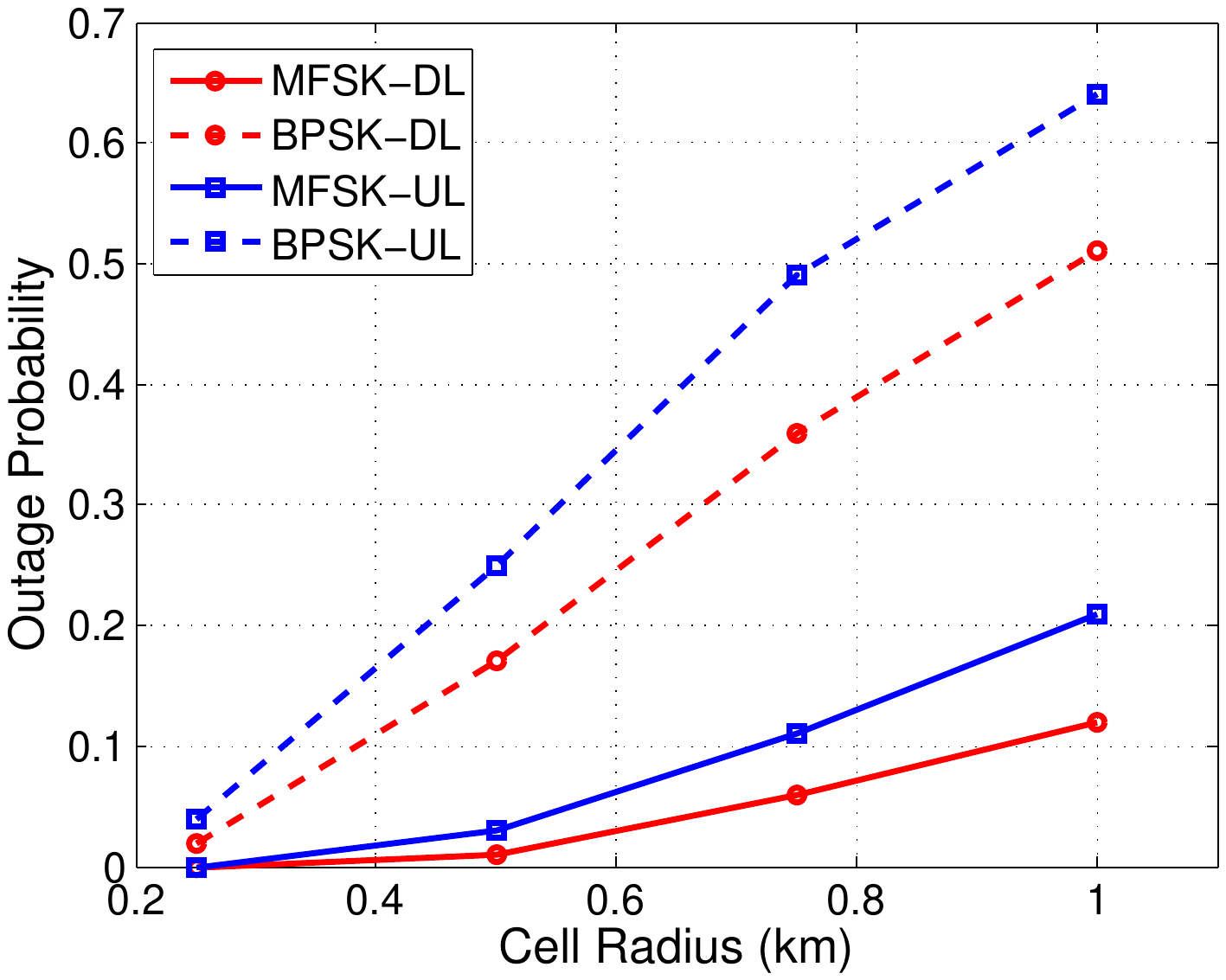}
\caption{Outage probability in the urban environment.}
\label{fig_8}
\end{figure}

\begin{figure}[hbpt!]
\centering
\includegraphics[trim=25mm 90mm 25mm 90mm,width=3.5in]{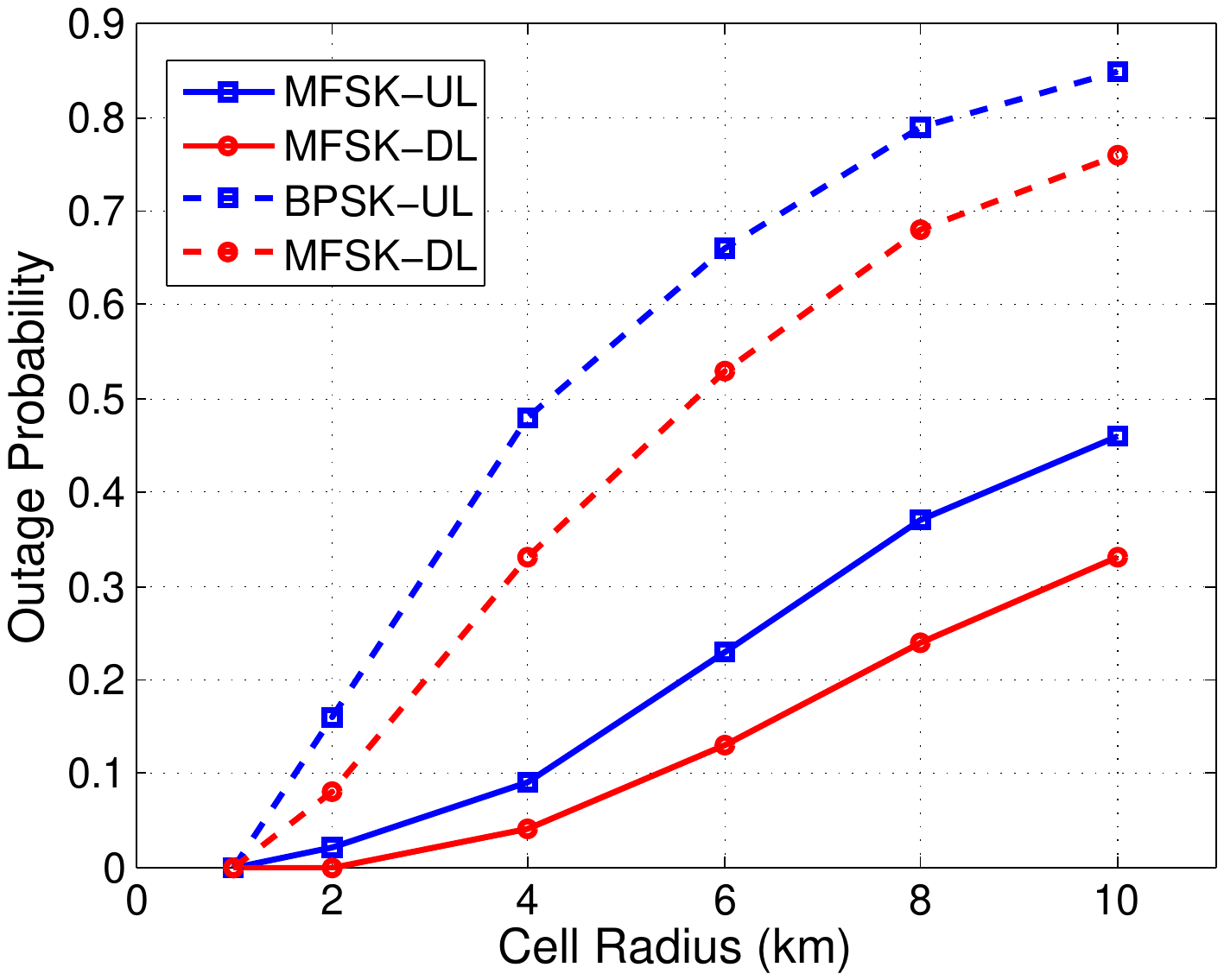}
\caption{Outage probability in the rural environment.}
\label{fig_9}
\end{figure}

\subsection{Capacity Analysis}
In this sub-section, we aim  to predict the aggregate throughput per sector and the available throughput per SM for the adaptive MFSK and compared the results with BPSK as shown in the following:
\subsubsection{Sector Capacity}
The average data rate per a sector can be evaluated based on the probabilities that each MFSK with a certain $M$ (i.e., MCS) is in use and their corresponding achievable data rate. The probability that MFSK with certain $M$ is in use can be calculated from the statistics of the received SNR in each scenario. The SNR values alter for different SMs due to different variables and assumptions such as the link distance, shadowing loss and the scenario type. The probability that the SNR values lie between the minimum SNR value desired by a given MCS and the SNR value desired by the next MCS is measured as follows:
\begin{equation}
    P_{MCS(i)}=Pr[\gamma_{o,MCS(i)}\leq SNR <\gamma_{o,MCS(i+1)}] 
\end{equation}

The total aggregate data rate (sector capacity) is obtained by applying the following relation:
\begin{equation}
    C_{total}=\sum_iC_{MCS(i)}P_{MCS(i)} ,
\end{equation}
where $C_{MCS(i)}$  is the data rate obtained when using the MCS(i) (i.e., MFSK with certain $M$ value). This result represents the upper band of the achievable throughput due to the assumptions such as data is always available to send and the actual throughput may be lower because of the retransmission process and under-utilisation  of the resource blocks \cite{RN66}.

\subsubsection{Available capacity per SM}
To predict the available capacity per SM, the minimum time interval between successive messages ($t_{min}$) and the average message size need to be identified. Based on \cite{RN66}, the SM message sizes for downlink \& uplink are selected to be 25 \& 2133 bytes, and 42 bytes as an overhead per each message is assumed.

To calculate $t_{min}$, the transport blocks number needed to send a message requires being determined. If $TBs_{i}$ is the transport block size for the MFSK with a certain $M$ (MCSi), then the transport block number needed to send a message with length $L$ bits equals:
 \begin{equation}
     N_{i}=\lceil \frac{message\,length(L)}{TBs_{i}}\rceil.
 \end{equation}
Then, the average of the total number of transport blocks ($N_{TB}$) required to the all SMs ($K$) in the sector  to send or receive a message can be expressed as \cite{RN66}:

\begin{equation}
    N_{TB}=K\sum_iP_{MCS(i)}N_{i}.
\end{equation}
Next, $t_{min}$ is determined as:
\begin{equation}
    t_{min}=\frac{N_{TB}}{R_{TB}},
\end{equation}
where, $R_{TB}$ is the rate of the transport block, which is equal to 21000 transport block per second in this paper (3MHz LTE-A like system is assumed). Ultimately, the upper bound of the available capacity per SM can be evaluated by dividing the message size to $t_{min}$.
\begin{figure}[hbpt!]
\centering
\includegraphics[trim=25mm 90mm 25mm 90mm,width=3.5in]{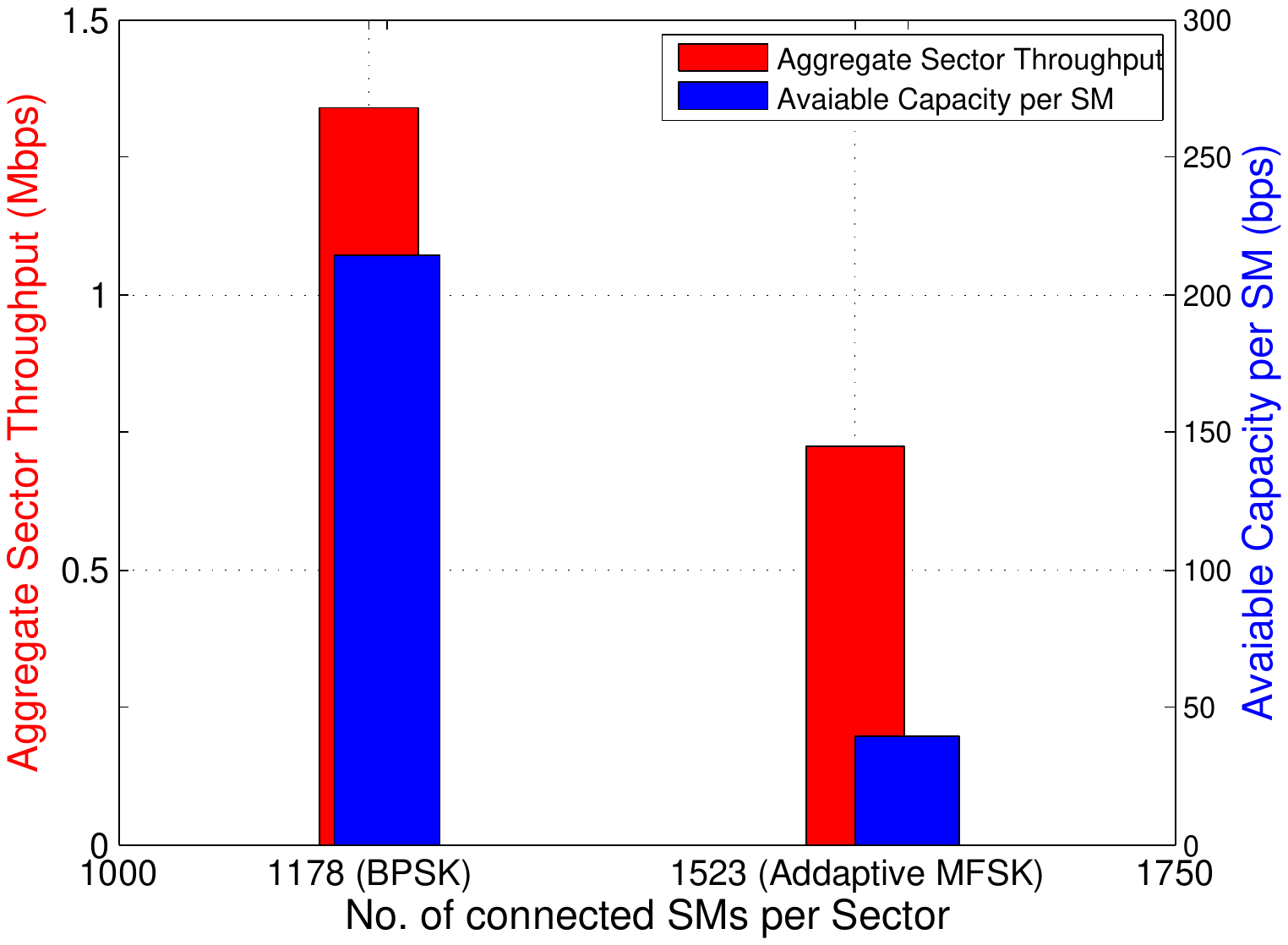}
\caption{Capacity analysis for the uplink urban case with R= 0.5 km.}
\label{fig_10}
\end{figure}

Fig. 10 \& 11 illustrate the aggregate sector throughput and the available capacity per SM versus the number of connected SMs per sector for 0.5 km urban \& 4 km rural uplink cases respectively. It is clear that the aggregate sector throughput and the available capacity per SM, in the adaptive MFSK case, decrease by around 46\% \& 81\% in the urban case and by 38\% \& 91\% in the rural case compared to the BPSK modulation scheme. However, the numbers of the connected SM per sector, in the adaptive MFSK, significantly increase by approximately 30\% \& 75\% in the urban and rural cases compared to BPSK. 

\begin{figure}[hbpt!]
\centering
\includegraphics[trim=25mm 90mm 25mm 90mm,width=3.5in]{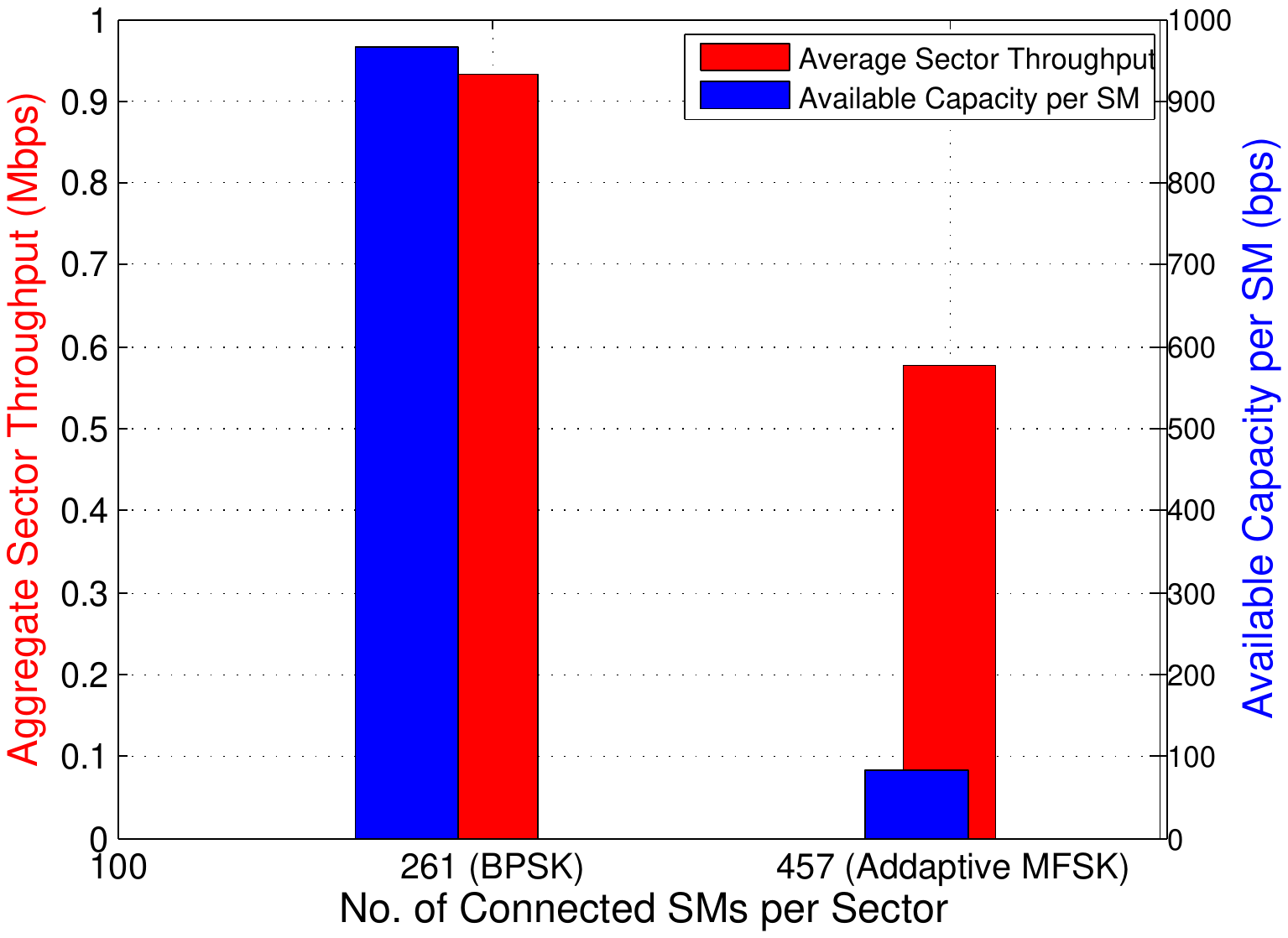}
\caption{Capacity analysis for the uplink rural case with R= 4 km.}
\label{fig_11}
\end{figure}

Tables-\ref{Table-IV}-\ref{Table-VII} show the aggregate sector throughput (Mbps), the available capacity per SM (bps) and the number of connected SMs per sector for uplink and downlink in the urban and rural cases respectively. Based on the results of all cases, the higher the cell radius, the lower the sector capacity and available capacity per SM. The reason for that is as the cell radius increases, a fraction of the cell area suffers from lower SNR increases. Finally, the difference in the sector throughput between both modulation schemes dramatically decreases as the cell radius increases for all cases.

\begin{table}[hbpt!]
\centering
\caption{Capacity Analysis for Uplink Urban Scenario.}
\label{Table-IV}
\begin{tabular}{|l|l|l|l|l|l|}
\hline
\multicolumn{2}{|l|}{\textbf{Cell radius (km)}} & 0.25 & 0.5 & 0.75 & 1 \\ \hline
\multirow{3}{*}{\textbf{MFSK}} & Agg. Sec Th.(Mbps) & 0.872 & 0.723 & 0.558 & 0.43 \\ \cline{2-6} 
 & Aval. Cap. per SM(bps) & 265.64 & 39.39 & 11.36 & 5.38 \\ \cline{2-6} 
 & No. of Connected SMs & 392 & 1523 & 3143 & 4961 \\ \hline
\multirow{3}{*}{\textbf{BPSK}} & Agg. Sec Th.(Mbps) & 1.72 & 1.34 & 0.91 & 0.65 \\ \cline{2-6} 
 & Aval. Cap. per SM(bps) & 669.6 & 214 & 139.9 & 111.46 \\ \cline{2-6} 
 & No. of Connected SMs & 376 & 1178 & 1801 & 2261 \\ \hline
\end{tabular}
\end{table}

\begin{table}[hbpt!]
\centering
\caption{Capacity Analysis for Downlink Urban Scenario.}
\label{Table-V}
\begin{tabular}{|l|l|c|c|c|c|}
\hline
\multicolumn{2}{|l|}{\textbf{Cell radius (Km)}} & 0.25 & 0.5 & 0.75 & 1 \\ \hline
\multirow{3}{*}{\textbf{MFSK}} & Agg. Sec Th.(Mbps) & 0.882 & 0.793 & 0.661 & 0.534 \\ \cline{2-6} 
 & Aval. Cap. per SM(bps) & 318.1 & 45.53 & 13.9 & 6.09 \\ \cline{2-6} 
 & No. of Connected SMs & 392 & 1554 & 3320 & 5526 \\ \hline
\multirow{3}{*}{\textbf{BPSK}} & Agg. Sec Th.(Mbps) & 1.76 & 1.49 & 1.15 & 0.88 \\ \cline{2-6} 
 & Aval. Cap. per SM(bps) & 651.11 & 191.95 & 110.65 & 81.29 \\ \cline{2-6} 
 & No. of Connected SMs & 384 & 1303 & 2260 & 3077 \\ \hline
\end{tabular}
\end{table}

\begin{table}[hbpt!]
\centering
\caption{Capacity Analysis for Uplink Rural Scenario.}
\label{Table-VI}
\begin{tabular}{|l|l|c|c|c|c|c|}
\hline
\multicolumn{2}{|l|}{\textbf{Cell radius (km)}} & 2 & 4 & 6 & 8 & 10 \\ \hline
\multirow{3}{*}{\textbf{MFSK}} & Agg. Sec Th.(Mbps) & 0.808 & 0.578 & 0.409 & 0.281 & 0.218 \\ \cline{2-7} 
 & Aval. Cap. per SM(bps) & 718 & 83.01 & 29.58 & 16.65 & 11.15 \\ \cline{2-7} 
 & No. of Connected SMs & 123 & 457 & 870 & 1266 & 1696 \\ \hline
\multirow{3}{*}{\textbf{BPSK}} & Agg. Sec Th.(Mbps) & 1.5 & 0.932 & 0.609 & 0.376 & 0.269 \\ \cline{2-7} 
 & Aval. Cap. per SM(bps) & 2400 & 965.4 & 655.9 & 597.31 & 536 \\ \cline{2-7} 
 & No. of Connected SMs & 105 & 261 & 384 & 422 & 471 \\ \hline
\end{tabular}
\end{table}

\begin{table}[hbpt!]
\centering
\caption{Capacity Analysis for Downlink Rural Scenario}
\label{Table-VII}
\begin{tabular}{|l|l|c|c|c|c|c|}
\hline
\multicolumn{2}{|l|}{\textbf{Cell radius (Km)}} & 2 & 4 & 6 & 8 & 10 \\ \hline
\multirow{3}{*}{\textbf{MFSK}} & Agg. Sec Th.(Mbps) & 0.85 & 0.683 & 0.519 & 0.384 & 0.305 \\ \cline{2-7} 
 & Aval. Cap. per SM(bps) & 707.5 & 98.3 & 32.4 & 15.82 & 10.29 \\ \cline{2-7} 
 & No. of Connected SMs & 125 & 482 & 983 & 1527 & 2104 \\ \hline
\multirow{3}{*}{\textbf{BPSK}} & Agg. Sec Th.(Mbps) & 1.69 & 1.2 & 0.84 & 0.57 & 0.43 \\ \cline{2-7} 
 & Aval. Cap. per SM(bps) & 2175 & 743.7 & 471 & 389.1 & 331.9 \\ \cline{2-7} 
 & No. of Connected SMs & 115 & 336 & 531 & 643 & 754 \\ \hline
\end{tabular}
\end{table}

\section{Conclusion}
In this paper, the OFDM-MFSK modulation scheme, which is based on the combination of OFDM and MFSK, is suggested for the smart metering technology. Its performance (PER, throughput, coverage and capacity) is measured and compared with the ordinary OFDM-BPSK in different cases. Based on the AWGN channel and system level study results, the OFDM-MFSK has better PER performance compared to BPSK at the higher values of $M$ ($M\geq8$), the higher the $M$ value, the better the PER performance. Whereas, for small $M$ values ($M<8$), the performance is worse than BPSK. On the other hand, the throughput behaviour is exactly the opposite to the PER behaviour. This is due to the fact that OFDM-MFSK is tailored to enhance the receiver sensitivity at the cost spectral efficiency. Additionally, the adaptive OFDM-MFSK has lower outage probability (higher coverage) compared to BPSK for both uplink and downlink in the urban and rural environments. Although the capacity for the adaptive OFDM-MFSK (aggregate sector throughput \& available capacity per SM) is lower than BPSK, the number of connected SMs per sector is higher. The essential requirements for the smart metering technology include the need for good coverage, low outage probability, and also the amount of data that is exchanged between the network and SMs is relatively low in this application. Therefore, we conclude that OFDM-MFSK can be effectively applied in the smart metering technology.


%


\section*{Acknowledgment}

Ghaith Al-Juboori would like to thank the Higher Committee for Education Development (HCED) in Iraq, Ministry of Oil and the University of Baghdad for sponsoring his Ph.D. studies.

\ifCLASSOPTIONcaptionsoff
  \newpage
\fi



%



\begin{IEEEbiography}[{\includegraphics[width=1in,height=1.25in,clip,keepaspectratio]{picture}}]{John Doe}
\blindtext
\end{IEEEbiography}

\bibliographystyle{IEEEtran}
\bibliography{IEEEabrv,References}

\end{document}